\documentclass[aps,twocolumn,amsmath,amssymb,floatfix,showpacs,
superscriptaddress]{revtex4}

\usepackage[dvips]{graphics}
\usepackage{color}
\definecolor{gold}{rgb}{0.85,0.66,0}
\definecolor{dblue}{rgb}{0,0,0.6}

\begin{document}

\title{Distribution of Persistent Currents in a Multi-Arm Mesoscopic Ring}

\author{Santanu K. Maiti}


\affiliation{Theoretical Condensed Matter Physics Division, Saha 
Institute of Nuclear Physics, Sector-I, Block-AF, Bidhannagar, 
Kolkata-700 064, India} 

\affiliation{Department of Physics, Narasinha Dutt College, 129 
Belilious Road, Howrah-711 101, India}

\author{Srilekha Saha}

\affiliation{Theoretical Condensed Matter Physics Division, Saha 
Institute of Nuclear Physics, Sector-I, Block-AF, Bidhannagar, 
Kolkata-700 064, India} 

\author {S. N. Karmakar}

\affiliation{Theoretical Condensed Matter Physics Division, Saha 
Institute of Nuclear Physics, Sector-I, Block-AF, Bidhannagar, 
Kolkata-700 064, India} 

\begin{abstract}
We propose an idea to investigate persistent current in individual 
arms of a multi-arm mesoscopic ring. Following a brief description 
of persistent current in a traditional Aharonov-Bohm (AB) ring, we 
examine the behavior of persistent currents in separate arms of a 
three-arm mesoscopic ring. Our analysis may be helpful in studying
magnetic response of any complicated quantum network. 
\end{abstract}


\maketitle

\section{Introduction}

Generation of persistent current in a normal metal mesoscopic ring 
threaded by an AB flux $\phi$ has been proposed over a number of decades. 
The appearance of discrete energy levels and large phase coherence length 
allow a non-decaying current in presence of an AB flux $\phi$. Following 
the pioneering work of B\"{u}ttiker {\em et al.}~\cite{butt}, various 
efforts have been made to explore the basic mechanisms of persistent 
current in mesoscopic rings and cylinders~\cite{cheu1,alts,schm,ambe,
san1,san2,belu,ore,peeters}. Later, the existence of non-decaying 
current in these systems has also been verified through some nice 
experiments~\cite{levy,chand,jari,deb}. The behavior of persistent current 
in a mesoscopic ring/cylinder can be studied theoretically by several ways 
as available in the literature~\cite{butt,cheu1,alts,schm,ambe,san1,san2,
belu,ore,peeters}. {\em In all these available procedures response 
of the entire system is achieved only, but no information about individual 
branches of the system can be explored though it is highly significant to 
elucidate the actual mechanism of electron transport in a more transparent 
way.} This is the main motivation behind this work.

With a brief description of persistent current in a single-channel 
mesoscopic ring, we discuss elaborately the behavior of persistent currents 
in individual arms of a three-arm mesoscopic ring. In this three-arm ring 
system we address an unconventional feature of persistent current when 
impurities are introduced only in the middle arm, keeping the other two 
arms free from any impurity. It shows that the current amplitude of the 
system increases with the increase of impurity strength in the strong 
impurity regime, while it decreases in the weak impurity regime. This 
phenomenon is completely different from traditional disordered systems,
and recently, few anomalous features of electron dynamics have also been 
reported in some other nano materials~\cite{zho1,ding1}.  

In what follows, we present the results. Section II is devoted to present 
the theory. The results are discussed in Section III. We provide a summary
in Section IV.

\section{Theoretical formulation}

\begin{figure}[ht]
{\centering \resizebox*{2.5cm}{2.5cm}{\includegraphics{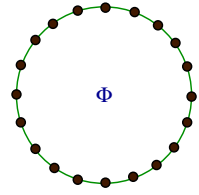}}\par}
\caption{A mesoscopic ring threaded by an AB flux $\phi$.}
\label{ring}
\end{figure}

\noindent
{\textbf{A simple ring:}} Let us start with Fig.~\ref{ring}, where a 
mesoscopic ring is subject to an AB flux $\phi$ (measured in unit 
of the elementary flux quantum $\phi_0=ch/e$). A tight-binding (TB) 
formalism is given for the description of the ring. Within the 
non-interacting picture the TB Hamiltonian for a $N$-site ring looks 
in the form,
\begin{equation}
\mbox{\boldmath $H$} = \sum_i \epsilon_i c_i^{\dagger} c_i + \sum_i v
\left(c_{i+1}^{\dagger}c_i e^{j\theta} + c_i^{\dagger}c_{i+1} 
e^{-j\theta} \right)
\label{equ1}
\end{equation}
where, $\epsilon_i$ is the site energy and $v$ gives the nearest-neighbor 
hopping integral. Due to the presence of AB flux $\phi$, a phase factor 
$\theta=2\pi\phi/N$ appears in the Hamiltonian. $c_i^{\dagger}$ ($c_i$) 
is the creation (annihilation) operator.

The energy $E_k$ corresponding to $k$-th energy eigenstate $|\psi_k\rangle$ 
can be obtained from the relation
$E_k=\langle \psi_k|\mbox{\boldmath $H$}| \psi_k \rangle$, where 
$|\psi_k\rangle = \sum\limits_p a_p |p\rangle$. Here $|p\rangle$'s are 
the Wannier states and $a_p$'s are the coefficients. Simplifying the
above relation we get the expression of the energy $E_k$ as,
\begin{equation}
E_k=\sum_i \epsilon_i a_i^* a_i + \sum_i v \left[a_{i+1}^* a_i 
e^{j\theta} + a_i^* a_{i+1} e^{-j\theta} \right]
\label{equ3}
\end{equation} 
where, $a_i^*$ is the complex conjugate of $a_i$. Here, the summation 
index $i$ runs over all the atomic sites in the ring.

Now, the current carried by the $k$-th energy eigenstate $|\psi_k\rangle$ 
can be determined from the relation,
\begin{eqnarray}
I_k = -\frac{\partial E_k}{\partial \phi} = \frac{2\pi j v}{N}\sum_i 
\left[a_i^* a_{i+1} e^{-j\theta}-a_{i+1}^* a_i e^{j\theta} \right].
\label{equ4}
\end{eqnarray}
At absolute zero temperature ($T=0$K), the net persistent current for a 
ring described with $N_e$ electrons can be determined by taking the sum 
of individual contributions from the lowest $N_e$ energy eigenstates. 
Therefore, for $N_e$ electron system the net current becomes
$I=\sum_{k=1}^{N_e} I_k$.

\begin{figure}[ht]
{\centering \resizebox*{3cm}{3cm}{\includegraphics{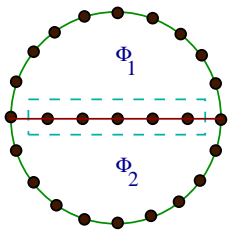}}\par}
\caption{A three-arm ring, where the upper and lower sub-rings are subject 
to AB fluxes $\phi_1$ and $\phi_2$, respectively.}
\label{ring1}
\end{figure}

\vskip 0.1cm
\noindent
{\textbf{A three-arm ring:}} 
Following the above prescription, the distribution of persistent currents
in individual arms, viz, upper, middle and lower, of a three-arm mesoscopic 
ring shown in Fig.~\ref{ring1} can be established. 

The TB Hamiltonian for this system is in the form,
\begin{eqnarray}
\mbox{\boldmath $H$}& =& \sum_i \epsilon_i c_i^{\dagger} c_i + \sum_i v_1
\left(c_{i+1}^{\dagger}c_i e^{j\theta_1} + c_i^{\dagger}c_{i+1} 
e^{-j\theta_1} \right) \nonumber \\
 & + & \sum_l \epsilon_l c_l^{\dagger} c_l + \sum_l v_2 
\left(c_{l+1}^{\dagger}c_l e^{j\theta_2} + c_l^{\dagger}c_{l+1} 
e^{-j\theta_2} \right)
\label{equ6}
\end{eqnarray}
where, the summation index $i$ is used to refer the atomic sites in the 
upper and lower arms of the ring i.e., in the outer ring, while the index 
$l$ describes the atomic sites in the middle arm (filled black circles in 
the framed region). $v_1$ and $v_2$ describe the nearest-neighbor hopping 
integrals in the outer ring and middle arm,
respectively. The phase factor $\theta_1$ is associated with the hopping 
of an electron in the upper/lower arm, while in the middle arm it is 
described by the term $\theta_2$. They are represented as,
$\theta_1=(\phi_1+\phi_2)/(N_U+N_L)$ and $\theta_2=(\phi_1-\phi_2)/2N_M$,
where $N_U$, $N_M$ and $N_L$ represent the total number of single bonds 
(a bond is formed by connecting two neighboring atoms through a line) in 
the upper, middle and lower arms of the ring, respectively. For this
geometry $E_k$ can be calculated according to the same prescription as
mentioned above, and, with this energy expression we can determine the 
currents in individual arms. The expressions for the currents are as follows.
\vskip 0.1cm
\noindent
For the upper arm:
\begin{equation}
I_k^U=\frac{2\pi j v_1}{N_U+N_L} \sum_i \left[a_i^* a_{i+1} 
e^{-j\theta_1}-a_{i+1}^* a_i e^{j\theta_1} \right]
\label{equ8}
\end{equation} 
where, contributions from the $N_U$ bonds are added. 
\vskip 0.1cm
\noindent
For the middle arm:
\begin{equation}
I_k^M=\frac{\pi j v_2}{N_M} \sum_l \left[a_l^* a_{l+1} 
e^{-j\theta_2}-a_{l+1}^* a_l e^{j\theta_2} \right]
\label{equ9}
\end{equation} 
here, net contribution comes from $N_M$ bonds.
\vskip 0.1cm
\noindent
For the lower arm:
\begin{equation}
I_k^L=\frac{2\pi j v_1}{N_U+N_L} \sum_i \left[a_i^* a_{i+1} 
e^{-j\theta_1}-a_{i+1}^* a_i e^{j\theta_1} \right]
\label{equ10}
\end{equation} 
here, the individual contributions of $N_L$ bonds are added. Using these 
relations (Eqs.~\ref{equ8}-\ref{equ10}), the net persistent currents at 
$T=0$K in individual arms of a three-arm mesoscopic ring containing $N_e$ 
electrons can be obtained in the same fashion as mentioned earlier.

In the present work we perform all the characteristics of persistent
current at $T=0$K and use the units where $c=h=e=1$. Throughout our 
numerical work we set the nearest-neighbor hopping strengths ($v$, $v_1$ 
and $v_2$) to $-1$ and the energy scale is measured in unit of $v$.

\section{Numerical results and discussion}

\begin{figure}[ht]
{\centering \resizebox*{7.5cm}{6cm}{\includegraphics{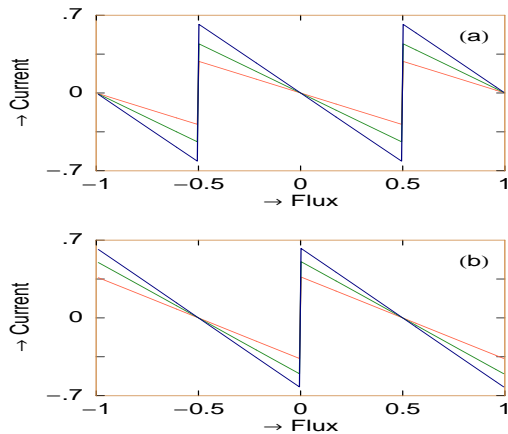}}\par}
\caption{$I$-$\phi$ spectra of an ordered mesoscopic ring ($N=20$) 
with (a) odd $N_e$ and (b) even $N_e$. The red, green and blue curves in 
(a) correspond to $N_e=3$, $5$ and $9$, respectively, while in (b) they 
represent $N_e=4$, $6$ and $10$, respectively.}
\label{current1}
\end{figure}

\noindent
{\textbf{A simple ring:}} As illustrative examples in Fig.~\ref{current1} 
we show the current-flux characteristics for a single-channel impurity-free 
($\epsilon_i=0$ for all $i$) mesoscopic ring considering $N=20$, where (a) 
and (b) correspond to the results for odd and even number of electrons, 
respectively. Both for the cases of odd and even $N_e$, current provides 
saw-tooth like nature with sharp jumps at half-integer or integer multiples  
of flux-quantum $\phi_0$ depending on whether the ring is described by odd 
or even number of electrons. These currents are periodic in $\phi$ 
exhibiting $\phi_0$ flux-quantum periodicity. The presented current-flux
characteristics for this single-channel ring exactly match with the 
previous theoretical studies where persistent currents have been calculated 
by other approaches~\cite{butt,cheu1,alts,schm,ambe,san1,san2,belu,
ore,peeters}.
 
Thus, making sure with the results for a single-channel ring now we can
safely use this procedure to illustrate the current-flux characteristics 
in individual arms of a three-arm mesoscopic ring.

\begin{figure}[ht]
{\centering \resizebox*{7.5cm}{6cm}{\includegraphics{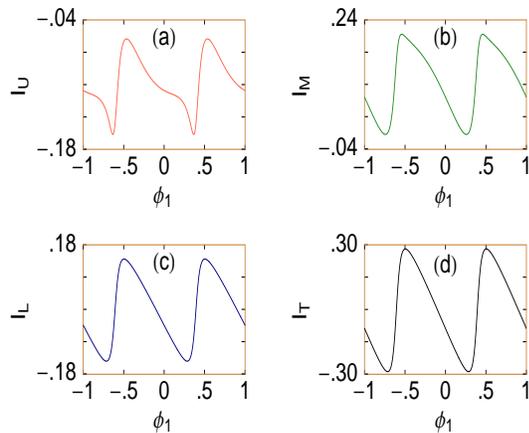}}\par}
\caption{Persistent current as a function of $\phi_1$ for an ordered 
three-arm mesoscopic ring considering $N_U=N_L=10$, $N_M=9$, $\phi_2=0.3$ 
and $N_e=10$. (a), (b) and (c) correspond to the currents in the upper, 
middle and lower arms, respectively, while (d) gives the total current 
for the entire three-arm ring.}
\label{current2}
\end{figure}

\vskip 0.1cm
\noindent
{\textbf{A three-arm ring:}} 
\vskip 0.1cm \noindent
{\em Ring without any impurity:}
In Fig.~\ref{current2} we present current-flux characteristics for an 
ordered ($\epsilon_i=\epsilon_l=0$ for all $i$ and $l$) three-arm 
mesoscopic ring, keeping $\phi_2$ constant. These currents are evaluated 
for a fixed number of electrons $N_e=10$ and they provide a complex spectra.
It is noticed that the responses of the individual branches are quite 
different from each other, especially, a significant change in amplitude 
takes place among the currents $I_U$, $I_M$ and $I_L$. By adding these 
three currents we get the net current ($I_T$) for the entire ring and it 
becomes exactly identical to the current determined from the other 
conventional methods available in the literature~\cite{butt,cheu1,
alts,schm,ambe,san1,san2,belu,ore,peeters}. This 
emphasizes the current conservation relation $I_T=I_U+I_M+I_L$. Other 
important feature is that the sum of the currents carried by the upper 
and middle arms becomes exactly identical to the current carried by the 
lower arm i.e., $I_L=I_U+I_M$ which gives another relation of current 
conservation. All these currents exhibit $\phi_0$ flux-quantum periodicity. 

Now, instead of $\phi_1$ if we plot the currents as a function of $\phi_2$, 
considering $\phi_1$ constant, exactly similar features are observed, like 
above, satisfying the current conservation relations. 

\vskip 0.1cm \noindent
{\em Ring with impurities in the middle arm:}
To introduce impurities in the middle arm we choose the site energies of 
the atomic sites, filled black circles in the framed region of 
Fig.~\ref{ring1}, randomly from a ``Box" distribution function of width 
$W$. Here, we determine the currents by taking the average over $50$ random 
disordered configuration in each case to achieve much accurate results.

\begin{figure}[ht]
{\centering \resizebox*{7.5cm}{8cm}{\includegraphics{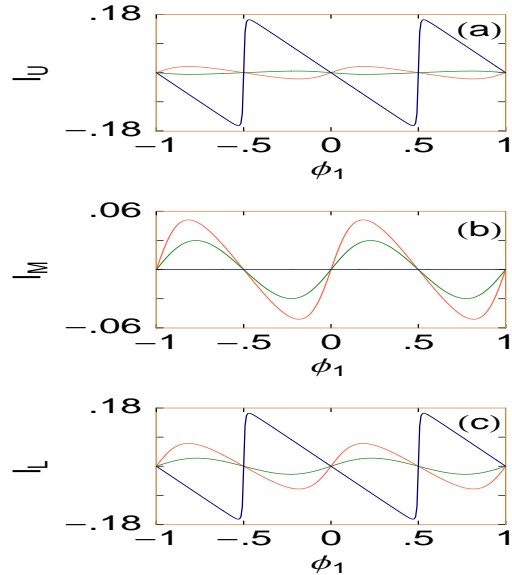}}\par}
\caption{Current-flux characteristics for the three different arms of a 
three-arm mesoscopic ring considering $N_U=N_L=15$, $N_M=11$, $\phi_2=0.5$ 
and $N_e=15$, where (a), (b) and (c) correspond to the currents in the upper, 
middle and lower arms, respectively. The red, green and blue curves in each 
panel represent $W=0$, $1$ and $10$, respectively.}
\label{current3}
\end{figure}

As the middle arm is subject to impurities, while the others are free, we
call this system as an ordered-disordered separated three-arm mesoscopic
ring. In such a mesoscopic ring an unconventional feature of persistent 
current is observed when we tune the strength of disorder $W$. To 
emphasize it, in Fig.~\ref{current3} we show the variation of persistent 
current in individual arms for a three-arm mesoscopic ring considering 
$N_U=N_L=15$ and $N_M=11$.
From the spectra it is observed that when $W=1$ (weak), the current 
amplitude gets reduced in all the three arms compared to the perfect case. 
The situation becomes completely different for the case of strong disorder 
i.e., $W=10$. In this case the current in the middle arm almost disappears, 
while for the other arms it achieves a high value. Thus a dramatic change 
in current amplitude takes place for the two different regimes of impurity 
strength. 

\begin{figure}[ht]
{\centering \resizebox*{7.5cm}{6cm}{\includegraphics{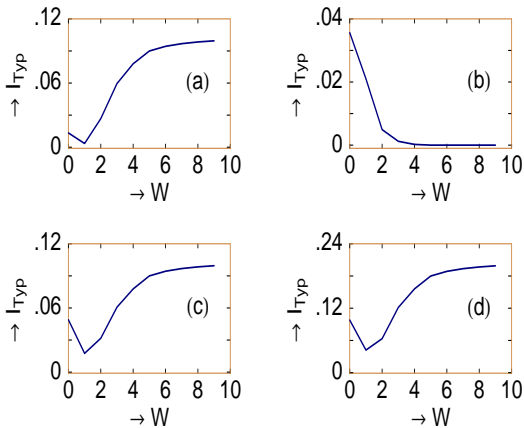}}\par}
\caption{$I_{\mbox{\scriptsize typ}}$-$W$ characteristics for a three-arm 
mesoscopic ring considering $N_U=N_L=15$, $N_M=11$, $\phi_2=0.5$ and 
$N_e=15$. (a), (b) and (c) represent the responses for the upper, middle 
and lower arms, respectively, while (d) corresponds to the variation for 
the entire ring.}
\label{typical}
\end{figure}

To explore this phenomenon more clearly, in Fig.~\ref{typical},
we show the variation of typical current amplitude as a function of $W$,
where (a), (b) and (c) give the results for the upper, middle and lower 
arms, respectively, and (d) corresponds to the variation for the entire 
ring. The typical current amplitude is evaluated from the relation,
$I_{\mbox{\scriptsize typ}}=\sqrt{\langle I^2 \rangle_{\phi_1,W}}$, 
where the averaging is done over a complete period of $\phi_1$ and $50$ 
random disordered configurations. From the $I_{\mbox{\scriptsize typ}}$-$W$ 
spectra the effect of disorder is clearly visible. 
It shows that the current amplitude in the middle arm sharply drops to 
zero as $W$ is increased. While, for the other two impurity-free arms the
current amplitude initially decreases, and reaching to a minimum at a 
critical value $W=W_c$ (say), it again increases with $W$. Thus, beyond 
the critical disorder strength $W_c$, the anomalous behavior is observed
and this phenomenon can be implemented as follows. We consider the 
three-arm mesoscopic ring as a coupled system combining two sub-systems,
one is ordered and other is disordered. In the absence of 
any coupling among the ordered and disordered regions, we can think the 
entire system as a simple combination of two independent sub-systems. 
Therefore, we get all the extended states in the ordered region, while 
the localized states are obtained in the disordered region. In this 
situation, the motion of an electron in any one region is not affected 
by the other. But for the coupled system, the motion of the electron is 
no more independent, and we have to take the combined effects coming from 
both the two regions. 
With the increase of disorder, the scattering effect becomes dominated 
more and thus the reduction of the current is expected. This scattering 
is due to the existence of the localized eigenstates in the disordered 
region. Now in the limit of weak disorder the coupling between the two 
sub-systems becomes strong and the motion of the electron in the ordered 
region is highly influenced by the disordered region. Therefore, the
scattering effect from both the two regions is quite significant and 
the current amplitude gets reduced. On the other hand, in the strong 
disorder regime the coupling between the two sub-systems becomes weak
and the scattering effect from the ordered region is less significant,
and it decreases with $W$. At the critical value $W_c$, we get a 
separation between the much weaker and the strongly localized states. 
Beyond this value, the weaker localized states become more extended and 
the strongly localized states become more localized with the increase 
of $W$. In this situation, the current is obtained mainly from these 
nearly extended states which provide the larger current with $W$ in 
the higher disorder regime.

From the above analysis, behavior of the typical current amplitude for 
the entire system (Fig.~\ref{typical}(d)) can be clearly understood
and it shows the similar variation like the upper and lower arms.

\section{Concluding remarks}

To summarize, we have explored an idea to investigate the nature of
persistent currents in individual branches of a multi-arm mesoscopic 
ring. Starting with a brief description of persistent current in a
traditional single-channel mesoscopic ring, pierced by an AB flux 
$\phi$, we have examined the characteristic features of persistent 
currents in separate arms of a three-arm mesoscopic ring where the 
upper and lower sub-rings are subject to AB fluxes $\phi_1$ and 
$\phi_2$, respectively. Our analysis may provide a basic framework 
to address magnetic response in individual branches of any complicated 
quantum network. 

In the present paper we have done all the calculations by ignoring the 
effects of temperature, electron-electron correlation, electron-phonon 
interaction, etc. We need further study by incorporating all these effects.

We are thankful to Prof. S. Sil, Prof. B. Bhattacharyya, M. Dey and 
P. Dutta for useful discussion.

\end{document}